\documentstyle[twocolumn,prl,aps,epsfig]{revtex}
\renewcommand{\vec}[1]{{\mathbf{#1}}}
\newcommand{\beq}{\begin{eqnarray}} 
\newcommand{\eeq}{\end{eqnarray}} 

\begin{document} 
\draft 
 
\title 
{Nearest-neighbour Attraction Stabilizes Staggered Currents in
the 2D Hubbard Model}
\author{Tudor D. Stanescu and Philip Phillips}

%
\address
{Loomis Laboratory of Physics\\
University of Illinois at Urbana-Champaign\\
1100 W.Green St., Urbana, IL, 61801-3080}

%

\address{\mbox{ }}
\address{\parbox{14.5cm}{\rm \mbox{ }\mbox{ }
Using a strong-coupling approach, we show that staggered
current vorticity does not obtain in the repulsive 2D Hubbard model 
for large on-site Coulomb interactions, as in the case of the copper
oxide superconductors.  This trend also persists even when nearest-neighbour
repulsions are present.  However, staggered
flux ordering emerges when
attractive nearest-neighbour
Coulomb interactions are included.  Such ordering opens a gap along
the $(\pi,0)-(0,\pi)$ direction
and persists over a reasonable range of doping. }}
\address{\mbox{ }}
\address{\mbox{ }}

\maketitle

Doped Mott insulators such as the cuprates are riddled with competing ordering
tendencies.  In addition to antiferromagnetism, $d_{x^2-y^2}$ pairing,
and charge ordering, staggered orbital antiferromagnetism produced by
local circulating
currents has joined the list of physical states that might occupy a central
place in the phase diagram of the cuprates.  Namely, staggered flux order\cite{chak}
might be at the heart of the elusive psuedogap phase.  It has been proposed
that such a state produces
a genuine gap in the single-particle spectrum at the $(\pi,0)$
point and
competes with d-wave pairing leading to the termination of superconductivity
in the underdoped regime.  Physically, currents
which alternate in sign from plaquette to plaquette in the copper-oxygen
plane comprise the staggered flux phase.
Consequently, the staggered flux phase can be thought of as
a density wave having $d_{x^2-y^2}$ symmetry. 
As charge density wave and superconducting order are well known to compete in
strongly-correlated systems, a psuedogap arising from a d-density wave state
has immediate appeal.

The earliest theoretical work\cite{marston} which showed that staggered
flux states might be the relevant low-energy excitations
of 2D Heisenberg antiferromagnets was based on $1/N$ expansions
of the $t-J$ model.  Such expansions cannot access the strong-coupling
limit relevant to the copper oxides.  Numerical simulations\cite{lee} using
a variational Gutzwiller-projected 
d-wave superconducting state for the
$t-J$ model
on a $10\times 10$ lattice revealed power law decay of the staggered current.
Leung\cite{leung} introduced two holes into a 32 site t-J cluster and also
found nonzero staggered vorticity correlations.  More recently,
density matrix renormalization group (DMRG) calculations\cite{white} on two-leg ladders
have found the rung-rung current correlation function to decay exponentially
indicating an absence of staggered flux ordering.  Consequently,
convincing computational evidence for staggered flux ordering is lacking.  In addition, 
there has been no study of the staggered flux phase based on a model
in which double occupancy is explicitly retained.  

To fill this void, we start with the Hubbard model.  We 
use the Hubbard operator
approach as these operators\cite{stanescu,compo} are tailor-made for treating
the strong-coupling regime, $U\gg t$ because they 
diagonalize the on-site Coulomb repulsion.
We have shown recently\cite{stanescu} that this approach yields a stable
$d_{x^2-y^2}$ superconducting state
in the absence of dynamical self-energy corrections.  We show here that
an equivalent treatment of the d-density wave state leads to
an absence of any long-range order of this type when $U$ is at least
on the order
of the bandwidth ($U=8t$), 
the relevant magnitude for the cuprates.  Absence of d-density wave ordering at this level 
of theory is significant because the dynamical or quantum corrections
tend to destroy long-range order.  Hence, we conclude that for $U=8t$,
d-density wave ordering is absent at any filling from the 2D Hubbard model.  
However, we find that when a nearest-neighbour
Coulomb interaction, $V=-0.25t$, is present, the staggered flux phase is stabilized.
A gap opens along the $(\pi,0)-(0,\pi)$ direction as
anticipated in Ref. (1).  Our results suggest that Coulomb attractions,
perhaps arising from phonons\cite{shen}, could be relevant to
the phenomenology of the cuprates.

The starting point for our analysis is the Hubbard model
\beq\label{HHam}
H = -\sum_{i,j,\sigma}t_{ij}c_{i\sigma}^{\dagger}c_{\sigma j} + 
U\sum_{i} n_{i\uparrow}n_{i\downarrow}+\sum_{i,j}V_{ij}n_in_j 
\eeq
with $t_{ij}=t\alpha_{ij}$ and $V_{ij}=V\alpha_{ij}$ where
$\alpha_{ij}=1$ when $ij$ are nearest-neighbours and zero otherwise
and $U$ the on-site Coulomb repulsion. 
For the cuprates,
$U\approx 8t$ and $t=0.5eV$.  
As $U$ is the largest energy scale in the problem, 
we use the eigen operators
of the atomic limit, namely 
$\eta_{i\sigma}=c_{i\sigma}n_{i-\sigma}$ and 
$\xi_{i\sigma}=c_{i\sigma}(1-n_{i-\sigma})$, which define the 
upper and lower Hubbard bands, respectively.
These operators 
diagonalise the on-site interaction term, $Un_{i\uparrow}n_{i\downarrow}=\frac{U}{2}
\sum_\sigma\eta_{i\sigma}^\dagger\eta_{i\sigma}$ and hence are the natural
starting point for a strong-coupling analysis.  The Hubbard operators are cumbersome
in that they do not obey usual fermionic anticommutation relationships. 
However, equation of motion techniques have proven quite useful
in circumventing
the lack of a diagrammatic scheme\cite{stanescu,compo}. We 
provide here only a sketch of the method as the full details
have been published elsewhere\cite{stanescu}.
First, we break the system into two sublattices and hence consider
the 4-component spinor,
\beq\label{psi}
\psi_\sigma(i,i')=\left(\begin{array}{l}
\psi^A_{i\sigma}\\\psi^B_{i'\sigma}
\end{array}\right)
\eeq
defined on two sites $i,i'$ with
\beq\label{psi}
\psi^A_\sigma(i)=\left(\begin{array}{l}
\xi^A_{i\sigma}\\\eta^A_{i\sigma}
\end{array}\right)\quad \psi^B_\sigma(i')=\left(\begin{array}{l}
\xi^B_{i'\sigma}\\\eta^B_{i'\sigma}
\end{array}\right).
\eeq
The key quantity of interest is the corresponding Green function,
\beq
G(i,i',t;j,j',t')&=& \theta(t-t')\langle\{\psi_\sigma(i,i',t),
\psi_\sigma(j,j',t'\}\rangle\nonumber\\
&=&\left(\begin{array}{ll}
G^{AA}(i,j)&G^{AB}(i,j')\\
G^{BA}(i',j)&G^{BB}(i',j')\end{array}\right).
\eeq
where $\{X,Y\}$ is the anticommutator and $\langle\cdots\rangle$ is the thermal
average.  Thermal averages involving 
operators on two sites $(i,j)$
\beq\label{cont}
\langle A_iB_j\rangle=\alpha_{ij}\langle AB'\rangle
+ie^{\vec Q\cdot\vec r_i}\gamma_{ij}\langle AB'\rangle_1
\eeq
will in general contain a real part, $\langle AB'\rangle$
and in the flux phase, an imaginary part, $\langle AB'\rangle_1$
as a consequence of the local broken time-reversal symmetry. 
Here, $\gamma_{ij}=1$ if
$ij$ are nearest neighbours along the x-axis, whereas $\gamma_{ij}=-1$
if the two neighbours are along the y-axis and zero otherwise,
$\vec Q=(\pi,\pi)$ so that the imaginary term alternates in sign
between the two sublattices
and the prime superscript signifies that the operators $A$ and 
$B$ are defined on nearest neighbour sites.  

To develop a system of self-consistent equations, we start by
writing the equations of motion,
\beq
j^{(1)}_{i\sigma}&=&i\dot{\xi}_{i\sigma}^A=-\mu\xi^A_{i\sigma}-\sum_j t_{ij}c_{j\sigma}^B-4t\pi_{i\sigma}^A+\xi_{i\sigma}^A\sum_j V_{ik} n_j^B\nonumber\\
j^{(2)}_{i\sigma}&=&-\dot{\eta}_{i\sigma}^A=-(\mu-U)\eta_{i\sigma}^A+4t\pi_{i\sigma}^A+\eta_{i\sigma}^A\sum_j V_{ij}n_j^B
\eeq
for the Hubbard operators in the presence of a finite chemical 
potential, $\mu$. 
The new operator $\pi$ is defined by
\beq
\pi_{i\sigma}^A=-\alpha_{ij}n_{i-\sigma}^Ac_{j\sigma}^B+\alpha_{ij}
c_{i-\sigma}^{A\dagger}c_{i\sigma}^Ac_{j-\sigma}^B
-\alpha_{ij}c_{i\sigma}^Ac_{i-\sigma}^Ac_{j-\sigma}^{A\dagger}
\eeq 
The remaining components, $j^{(3)}$ and $j^{(4)}$
are constructed simply by interchanging $A$ and $B$ and $i$ and $i'$ in
$j^{(1)}$ and $j^{(2)}$, respectively.  In the static (or pole)
approximation\cite{compo}, the Green
function equations
are closed using the Roth\cite{roth} projector
\beq
{\cal P}(j_i)=\sum_{ln}\langle\{j_i,\psi_l^\dagger\}\rangle I_{ln}^{-1}\psi_n.
\eeq
which explicitly removes the components of $j^{(i)}$ which are 
orthogonal to the Hubbard basis.  The overlap
matrix ${\bf I}=F.T.\langle\{\psi_\sigma(i,i'),\psi_\sigma^\dagger(j,j')\}\rangle$
is an enitirely diagonal $4\times 4$ matrix with 
$I_{11}=I_{33}=1-\langle n_{-\sigma}\rangle$ and 
$I_{22}=I_{44}=\langle n_{-\sigma}\rangle$. Here $F.T.$ signifies the
Fourier transform.
At this
level of theory, the energy levels are sharp as the Green function
\beq\label{gfeq}
G(\vec k,\omega)=(\omega{\bf 1}-\vec E)^{-1}I=\sum_{i=1}^4\frac{\sigma_i(\vec k)}{\omega-E_i(\vec k)+i\eta}
\eeq
has poles at the real energies, $E_i(k)$, where $i=1,2$ indexes the
lower Hubbard band and $i=3,4$ the upper Hubbard band.  Because of 
the two sublattice construction, $E_1(\vec k)=E_2(\vec k+\vec Q)$
and likewise, $E_3(\vec k)=E_4(\vec k+\vec Q)$. In
Eq. (\ref{gfeq}), $\sigma_i=z(E_i)/\prod_{j\ne i}(E_i-E_j)$ 
where $z(\omega)=
{\rm Det}(\omega {\bf 1}-{\bf E})(\omega 
{\bf 1}-{\bf E})^{-1}{\bf I}$.  The poles of the 
Green function result from diagonalising the matrix
\beq
\vec E=\vec M\vec I^{-1}
\eeq
where 
and $\vec M=F.T.\langle\{j_\sigma(i,i'),\psi^\dagger_\sigma(i,j')\}\rangle$ 
is a $4\times 4$ matrix defined on the two sublattices.  The matrix elements of $\vec M$ contain 
two types of correlation functions.  The first class, 
$a=\langle\xi_{i\sigma}^A\xi_{j\sigma}^{B\dagger}\rangle$, $b=\langle \xi_{i\sigma}^A\eta_{j\sigma}^{B\dagger}\rangle=\langle\eta_{i\sigma}^A\xi_{j\sigma}^{B\dagger}\rangle$,
and $c=\langle\eta^A_{i\sigma}\eta_{j\sigma}^{B\dagger}\rangle$
involves operators which appear in the Hubbard basis and hence enter directly as self-consistent parameters in the closure for the Green function.
In the second class, correlations of the form,
\beq
p=\langle n_{i\sigma}^A n_{j\sigma}^B\rangle+\langle c_{i\sigma}^{A\dagger}c_{i-\sigma}^Ac_{j-\sigma}^{B\dagger}c_{j\sigma}^B\rangle-
\langle c_{i\sigma}^Ac_{i-\sigma}^Ac_{j-\sigma}^{B\dagger}c_{j\sigma}^{B\dagger}\rangle
\eeq
which corresponds to the bandwidth renormalization
and $g={\rm Re}\langle n_{i\sigma}^A n_{j\sigma}^B\rangle$,  arise between
entities outside the Hubbard basis.  Both of these correlation functions must be
decoupled and expressed in terms of correlation functions of the type $a-c$. 
Expressions appear in our earlier paper\cite{stanescu} for the real part of 
the correlation function $p$.  The important difference here is that now
$p$ is complex.  Its imaginary part is given by
\beq\label{imp1}
{\rm Im} p\equiv p_1=\frac{\delta_1}{\alpha+1}
\eeq
where
\beq
\alpha=\frac{1}{\langle n_{-\sigma}\rangle}\left(C_{22}+C_{12}\right)
-\frac{1}{1-\langle n_{-\sigma}\rangle}\left(C_{11}+C_{21}\right)
\eeq
and
\beq
\delta_1&=&\frac{4}{\langle n_{-\sigma}\rangle(2-\langle n_{-\sigma}\rangle)}
\left[{\rm Re}(a+b){\rm Im}(c+b)\right.\nonumber\\
&&\left.+{\rm Im}(a+b){\rm Re}(c+b)\right]
\eeq
where $C_{12}=\langle\xi_{i\sigma}\eta_{i\sigma}^\dagger\rangle=C_{21}$,
$C_{11}=\langle\xi_{i\sigma}\xi_{i\sigma}^\dagger\rangle$,
and $C_{22}=\langle\eta_{i\sigma}\eta_{i\sigma}^\dagger\rangle$.

\begin{figure}
\begin{center}
\epsfig{file=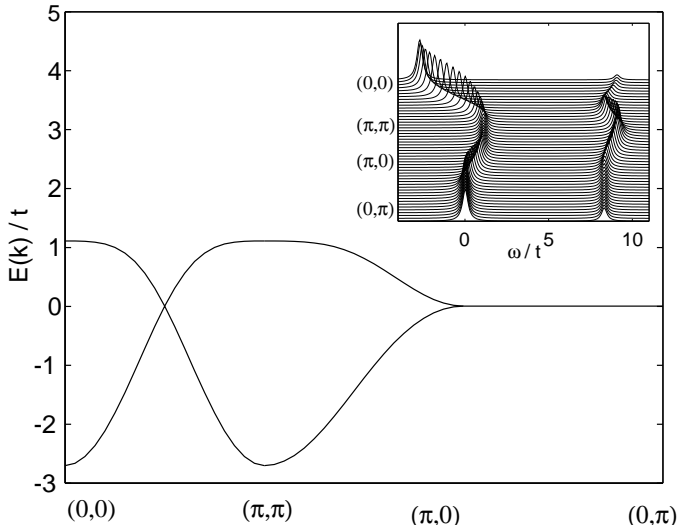, height=7cm}
\caption{Momentum dependence of the lower Hubbard band for
$U=8t$, $n=0.79$, and $T=0.01t$.  From the spectral function
shown in the inset, it clear that the spectral weight in the lower Hubbard
band resides entirely on the curve with a flat dispersion
near $(\pi,\pi)$. The spectral weight in the upper Hubbard band is also shown.}
\label{fig1}
\end{center}
\end{figure}

Noting that any correlation function of the form, 
$\langle\psi_m(i)\psi^\dagger_n(j)\rangle$ is related to the Green function
through
\beq\label{corr1}
\langle\psi_m(i)\psi^\dagger_n(j)\rangle&=&\frac{\Omega}{(2\pi)^2}\int d^2k d\omega
e^{i\vec k\cdot(\vec r_i-\vec r_j)}(1-f(\omega))\nonumber\\
&&\times \sum_{i=1}^4\left(\sigma_i\right)_{mn}(\vec k)
\delta(\omega-E_i(\vec k))
\eeq
the complete set of equations for the correlation
functions, $a$, $b$, $c$, $p$, and $g$ together with
the equation for the Green function, Eq. (\ref{gfeq}), represent a closed
set of integral equations that must be solved self-consistently.
In the absence of nearest-neighbour interactions, the terms
containing $a$, $b$, and $c$ are summed over nearest neighbour
sites.  Hence, the imaginary part cancels as a result of the 
alternating signs.  Such is not the case for $p$, however. 
Consequently, staggered current vorticity
is stable if the closed set of equations admits
a non-zero value of ${\rm Im}p=p_1$, the key signature
of the broken symmetry.
Shown in Fig. (\ref{fig1}) is the self-consistent solution
for the lower Hubbard bands for $U=8t$, $n=0.79$, $T=0.01t$ and 
$V=0$. 
The two bands in Fig. (\ref{fig1}) correspond to $E_1(\vec k)$
and $E_2(\vec k)=E_1(\vec k+Q)$.
As the spectral
function attests (see inset Fig. (1)), only one of the bands has a
non-zero weight as is expected in the absence of staggered currents.
The spectral function shown in the inset
$-1/\pi{\rm Im}\sum_{mn}\left(G^{AA}+G^{AB}\right)_{mn}$ in actuality consists 
of a series of $\delta$-functions.
Each peak has been given an artificial width of $0.2t$.

\begin{figure}
\begin{center}
\epsfig{file=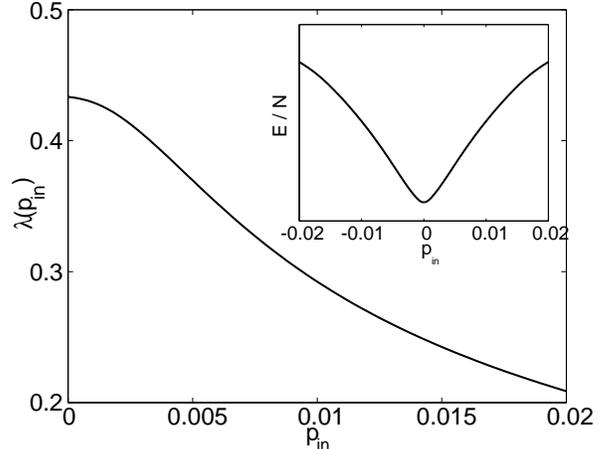, height=6cm}
\caption{Staggered flux stability parameter $\lambda$ vs 
the input value ($p_{\rm in}$) for the imaginary part of the correlation
function, $p$.   The flux phase is 
stable if $\lambda=1$ and $d\lambda/\d p_{\rm in}<0$.  Because
$\lambda<1$, staggered currents are absent.  The inset demonstrates
that the energy per particle is a minimum when 
$p$ is real, thereby implying an absence of the staggered current
phase.  Here
$U=8t$ and $V=0$. }
\label{fig2}
\end{center}
\end{figure}

To investigate the stability of staggered currents, we set $p_1$ to a fixed
value, $p_{\rm in}$, and determine the self-consistent values of all 
correlation functions upon successive iteration.  We then obtained
the output value of $p_1=p_{\rm out}$ by using
Eq. (\ref{imp1}). Staggered current vorticity is stable
if $p_{\rm in}=p_{\rm out}$ or equivalently,
$\lambda(p_{\rm in})=p_{\rm out}/p_{\rm in}=1$ 
and $d\lambda/dp_{\rm in}<0$.
As is evident from Fig. (\ref{fig2}), $\lambda<1$ and
it is
a decreasing function of $p_{\rm in}$, proving that staggered
currents are absent for $U=8t$, $V=0.0$ and $T=0.01t$.  This can also be seen
from the inset of Fig. (\ref{fig2}) which
contains the average energy per particle ($\langle H\rangle$) as
a function of $p_1$.  The minimum
value of the energy corresponds to $p_1=0$, further affirming the absence
of staggered currents.

\begin{figure}
\begin{center}
\epsfig{file=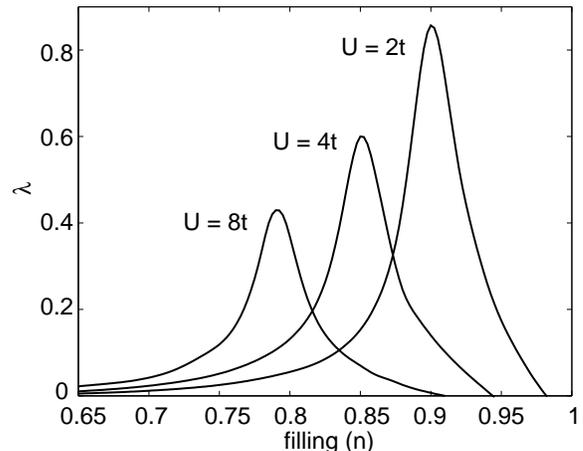, height=6cm}
\caption{Density dependence of the staggered current stability parameter.
Half-filling corresponds to $n=1$.}
\label{fig3}
\end{center}
\end{figure}

Although the flux phase is absent for the parameters above, we can examine the 
tendency towards
such ordering as a function of density for a sufficiently small
value of $p_1$ such that $\lambda\approx \lambda (0)$, the maximum
value of $\lambda$ as shown in Fig. (\ref{fig2}). 
Shown in Fig. (\ref{fig3}) is the density dependence of the tendency
towards staggered flux ordering 
for $p_1=0.001$ for three values of the on-site Coulomb repulsion,
$U=8t,4t,2t$, $V=0.0$ 
and $T=0.01t$. 
As is evident, stable staggered currents are most likely
at small values of $U$ and as the density approaches half-filling as has been
 reported earlier in mean-field studies
of the $t-J$ model\cite{marston}.  However, for the parameters of the Hubbard
model that are relevant for the copper oxides, $U\gg t$, such ordering 
is absent at this level of theory.  Going beyond the static
approximation involves including quantum fluctuations.  Such processes
can only suppress ordering.  Hence, we conclude that for $U\gg t$, the
Hubbard model (with $V=0$) cannot sustain staggered flux ordering.

\begin{figure}
\begin{center}
\epsfig{file=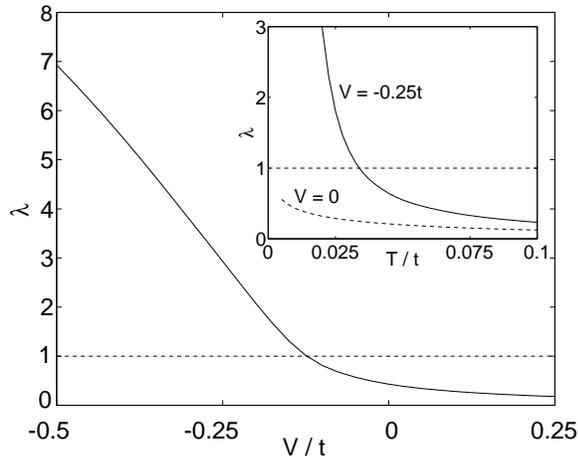, height=6cm}
\caption{Staggered current stability parameter as a function of the 
nearest-neighbour
Coulomb interaction, $V$ for $n=0.79$, $U=8t$ and $T=0.01t$.  
For $V<0$, the stability parameter exceeds unity
leading to an onset of staggered flux order.  From the inset, the onset temperature
when $V=-0.25t$ is $T=0.33t\approx 160K$.}
\label{fig4}
\end{center}
\end{figure}
\begin{figure}
\begin{center}
\epsfig{file=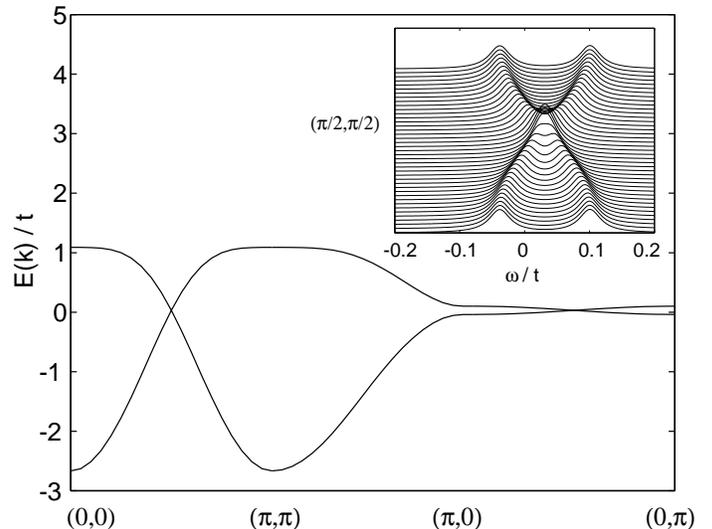, height=7cm}
\caption{Lower Hubbard energy bands and spectral function 
for $n=0.79$, $V=-0.25t$ and
$U=8t$.  The presence of a gap along the $(\pi,0)-(0,\pi)$ direction
with a node at $(\pi/2,\pi/2)$ is indicative of 
$d_{x^2-y^2}$ order.}
\label{fig5}
\end{center}
\end{figure}

What about nearest-neighbour Coulomb interactions?  As is evident from
Fig. (\ref{fig4}), the flux phase is destabilised even further for $V>0$.
However, for $V<0$, $\lambda$ crosses unity and staggered flux
ordering obtains. This is the primary result of
this study.
The inset shows the temperature dependence of 
$\lambda$ for $V=0.0$ as well as for $V=-0.25t$.  The onset temperature
for the flux phase is $T\approx 0.033t\approx 160K$. For $V=0.0$,
$\lambda$ is relatively flat indicating further that even
at $T=0$, such ordering is absent.   To examine the impact of the 
staggered current on the band structure, we recalculated the lower Hubbard band
for $V=-0.25t$, $n=0.79$, $U=8t$ and $T=0.01t$.  From Fig. (\ref{fig5}), 
we see that
a non-zero staggered current vorticity has opened a gap near the Fermi energy
along the $(\pi,0)-(0,\pi)$ direction. 
The inset shows more clearly
that the Fermi level lies in the gap near the $(\pi,0)$ and $(0,\pi)$
regions.  The position of the Fermi
energy relative to the gap edge is doping dependent.  

Our work clearly shows that for the parameters relevant
to the copper oxides, staggered current vorticity cannot be 
stabilized unless attractive interactions are present.  Further, the apparent
doping dependence (see Fig. (\ref{fig3})) of the staggered current 
is not entirely consistent with competing order in the underdoped regime. 
Hence, either staggered flux order is irrelevant to the copper oxides or at least
nearest neighbour attractions are {\bf essential} to their stabilization.  
An obvious source of such an attraction are local phonon modes as
has been discussed recently\cite{shen}.  Our work suggests (assuming
the flux phase is relevant) that
the strange phenomenology of the cuprates arises from cooperation between
phonons and strong on-site Coulomb repulsions.  
\acknowledgements
We thank Brad Marston for a useful e-mail exchange and
the NSF grant No. DMR98-96134.

\end{document}